\documentclass[11pt,a4paper,english,superscriptaddress,aps,preprint]{revtex4}
\usepackage{xcolor}
\usepackage{amsfonts}
\usepackage{graphics}
\usepackage{graphicx}
\usepackage{hyperref}
\usepackage{slashed}
\usepackage{amsmath}
\usepackage{amssymb}
\makeatletter
\usepackage{babel}
\newcommand{\bea}{\begin{eqnarray}}
\newcommand{\eea}{\end{eqnarray}}

\begin{document}

\title{One-loop effective potential in the non-local supersymmetric gauge theory}

\author{F. S. Gama}
\email{fisicofabricio@yahoo.com.br}
\affiliation{Departamento de F\'{\i}sica, Universidade Federal da Para\'{\i}ba\\
 Caixa Postal 5008, 58051-970, Jo\~ao Pessoa, Para\'{\i}ba, Brazil}

\author{J. R. Nascimento}
\email{jroberto@fisica.ufpb.br}
\affiliation{Departamento de F\'{\i}sica, Universidade Federal da Para\'{\i}ba\\
 Caixa Postal 5008, 58051-970, Jo\~ao Pessoa, Para\'{\i}ba, Brazil}

\author{A. Yu. Petrov}
\email{petrov@fisica.ufpb.br}
\affiliation{Departamento de F\'{\i}sica, Universidade Federal da Para\'{\i}ba\\
 Caixa Postal 5008, 58051-970, Jo\~ao Pessoa, Para\'{\i}ba, Brazil}

\author{P. J. Porf\'{i}rio}
\email{pporfirio89@gmail.com}
\affiliation{Departamento de F\'{\i}sica, Universidade Federal da Para\'{\i}ba\\
 Caixa Postal 5008, 58051-970, Jo\~ao Pessoa, Para\'{\i}ba, Brazil}

\begin{abstract}
Using the superfield formalism, we propose a non-local extension of the supersymmetric gauge theory coupled to massless chiral matter.  Two different non-local models are considered. For these models, we explicitly calculate the one-loop K\"{a}lerian effective potential.
\end{abstract}

\maketitle

\section{Introduction}

In the recent years, nonlocal extensions of field theory models began to be intensively studied. There are two main reason for it: first, the nonlocality naturally arises within studies of finite-size objects (for general field theory aspects and phenomenological applications, see \cite{Efimov}, and for the discussion of nonlocality within the string context, see f.e. \cite{CalMod}), second, the nonlocal approach appears to be a very convenient and powerful aim for consistent solutions of fundamental problems of quantum gravity. Indeed, it is well known that the Einstein gravity is non-renormalizable, while its higher-derivative extensions are known to involve undesirable ghost-like excitations \cite{Ant}. At the same time, an appropriate nonlocal extension of gravity allows to solve both these problems, because, first, of a very good ultraviolet asymptotics of corresponding propagators, second, of its specific momentum behavior represented by a so-called entire function which does not allow for arising the ghosts \cite{ModRachw}. This clearly calls the interest to quantum properties of field theory models including the gravity. 

Moreover,  nonlocal extensions for different field theory models clearly can be treated as a convenient laboratory for study of many issues related to nonlocality, keeping in the mind that the final aim of this research line, that is, quantum calculations in possible nonlocal extensions of gravity, is a very complicated problem. Thus, loop calculations of various nonlocal field theory models naturally attracts the interest. An important step along this line has been done in the paper \cite{Briscese} where the one-loop effective potential for different nonlocal scalar field models has been calculated, with the characteristic nonlocality scale turns out to play the role of the UV regularization parameter, i.e. a nonlocal extension represents itself as a specific form of the higher-derivative regularization. Further, this methodology has been generalized for supersymmetric field theories formulated with use of the superfield approach, so, in \cite{BGNP}. one-loop low-energy effective action has been calculated for different nonlocal extensions of the Wess-Zumino model. Because of the high and deeply motivated interest to supergauge theories, the natural development of this study consists in introducing the nonlocal extension for the ${\cal N}=1$ super-Yang-Mills theory with chiral matter with the subsequent calculation of the one-loop K\"{a}hlerian effective potential which is known to characterize the low-energy effective action. This is the aim we pursue in this paper. Our calculations are essentially based on the superfield approach.
Throughout this work, we follow the conventions adopted in \cite{GGRS}. 

The structure of the paper looks like follows: in the section 2, we introduce an action of the supersymmetric non-Abelian gauge theory, in the section 3, the general expression for the one-loop K\"{a}hlerian effective potential in this theory is presented, and in the section 4, it is explicitly evaluated for different choices of nonlocal form factors. The section 5 summarizes the results.


\section{Non-local Supersymmetric Gauge Theory}

Let us start our study with a higher-derivative version of the supersymmetric non-Abelian gauge theory. We define the classical action for this theory in the $\mathcal{N}=1$, $d=4$ superspace as: 
\bea
\label{gaugeaction}
S_{G}[V]=\frac{1}{16g^2}\textrm{Tr}\left\{\int d^6z W^\alpha f\left(\Box_c\right)W_\alpha+\int d^6\bar z \bar W^{\dot{\alpha}}f\left(\Box_a\right)\bar W_{\dot{\alpha}}\right\} \ ,
\eea
where $S_{G}$ is a functional of a Lie-algebra-valued scalar superfield $V^AT_A$ ($T_A$ are the Lie algebra generators in the adjoint representation of the gauge group). Here, the field strength spinors are given in terms of the gauge superfield $V(z)$ by 
\bea
\label{strength}
W_\alpha=i\bar D^2\left(e^{-2gV}D_\alpha e^{2gV}\right) \ , \ \bar W_{\dot{\alpha}}=iD^2\left(e^{2gV}\bar D_{\dot{\alpha}} e^{-2gV}\right)  \ .
\eea
We assume that the dimensionless kinetic operators $f(\Box_c)$ and $f(\Box_a)$ coincide with the identity in some suitable limit. Additionally, we also assume that $f(\Box_c)$ and $f(\Box_a)$ can be represented by infinite series of the chiral covariant d'Alembertian $\Box_c$ and the antichiral one $\Box_a$, respectively \cite{Efimov}. These d'Alembertian operators are defined by \cite{KM}
\bea
\Box_c&=&\Box_{cov}^{(+)}-iW^\alpha\nabla_\alpha^{(+)}-\frac{i}{2}\left(\nabla^{(+)\alpha} W_\alpha\right) \ , \ \Box_c W_\alpha=\bar\nabla^{(+)2}\nabla^{(+)2} W_\alpha \ , \ \bar\nabla^{(+)}_{\dot{\alpha}} W_\alpha=0 \ ;\\
\Box_a&=&\Box_{cov}^{(-)}-i\bar W^{\dot{\alpha}}\bar\nabla_{\dot{\alpha}}^{(-)}-\frac{i}{2}\left(\bar\nabla^{(-)\dot{\alpha}}\bar W_{\dot{\alpha}}\right) \ , \ \Box_a \bar W_{\dot{\alpha}}=\nabla^{(-)2}\bar\nabla^{(-)2} \bar W_{\dot{\alpha}} \ , \ \nabla^{(-)}_\alpha \bar W_{\dot{\alpha}}=0 \ ,
\eea
where $\Box_{cov}=\frac{1}{2}\nabla^{\alpha\dot{\alpha}}\nabla_{\alpha\dot{\alpha}}$. The superscripts $(+)$ and $(-)$ indicate that the above derivatives are gauge covariant with respect to $\Lambda$- and $\bar\Lambda$-parameters, where $\Lambda$ and $\bar\Lambda$ are chiral and antichiral parameters, respectively \cite{BK}. Thus, the gauge covariance with respect to $\Lambda$- and $\bar\Lambda$-parameters requires that the derivatives be defined in terms of $V(z)$ as \cite{GGRS}
\bea
\label{representation}
\left(\nabla_\alpha^{(+)} \ , \ \bar\nabla_{\dot{\alpha}}^{(+)} \ , \ \nabla_{\alpha\dot{\alpha}}^{(+)}\right)&=&\left(e^{-2gV}D_\alpha e^{2gV} \ , \ \bar D_{\dot{\alpha}} \ , \ -i\left\{\nabla_\alpha^{(+)} \ , \ \bar\nabla_{\dot{\alpha}}^{(+)}\right\}\right) \ ; \\
\left(\nabla_\alpha^{(-)} \ , \ \bar\nabla_{\dot{\alpha}}^{(-)} \ , \ \nabla_{\alpha\dot{\alpha}}^{(-)}\right)&=&\left(D_\alpha \ , \  e^{2gV}\bar D_{\dot{\alpha}}e^{-2gV} \ , \ -i\left\{\nabla_\alpha^{(-)} \ , \ \bar\nabla_{\dot{\alpha}}^{(-)}\right\}\right) \ .
\eea 
With the definitions above, it is important to note that (\ref{gaugeaction}) is invariant under the gauge transformation $e^{2gV}\rightarrow e^{2ig\bar\Lambda}e^{2gV}e^{-2ig\Lambda}$.

It is well known that higher-derivative theories tend to contain extra degrees of freedom with negative energy, which lead to a Hamiltonian that is bounded neither from below nor from above \cite{Ghost}. Thus, in order to avoid this problem in our model (\ref{gaugeaction}), we must impose an additional constraint on the kinetic operators $f(\Box_c)$ and $f(\Box_a)$. This additional constraint can be determined through the pole structure of the propagator. Therefore, let us calculate the propagator for the gauge superfield $V(z)$. For this, we must first add to (\ref{gaugeaction}) a gauge-fixing term. In this work, we will consider following higher-derivative generalization of the usual gauge-fixing term:
\bea
\label{gaugefixing}
S_{GF}[V]=-\frac{1}{\alpha}\textrm{Tr}\int d^8z\left(g\left(\Box\right)D^2V\right)\left(g\left(\Box\right)\bar D^2V\right) \ .
\eea
It follows from (\ref{gaugeaction}) and (\ref{gaugefixing}) that the propagator for the gauge superfield $V\left(z\right)$ is given by
\bea
\label{prop}
\left\langle V^A\left(-p,\theta\right)V^B\left(p,\theta^\prime\right)\right\rangle=-\frac{1}{p^4}\left[\frac{1}{f\left(-p^2\right)}D^\alpha\bar D^2D_\alpha-\frac{\alpha}{g^2\left(-p^2\right)}\left\{D^2,\bar D^2\right\}\right]\delta^{AB}\delta^4(\theta-\theta^\prime) \ .
\eea
Thus, in order to ensure that the theory (\ref{gaugeaction}) does not contain new degrees of freedom as compared to the standard gauge theory, we must require that $f(-p^2)$ is an entire function with no zeros and $f(-p^2)=g^2(-p^2)$  \cite{Efimov,BMS}. Based on these requirements, we can state that $f(-p^2)=g^2(-p^2)=e^{-h(-p^2)}$, where $h(-p^2)$ is an entire function \cite{APP} which we will call the form factor. Therefore, we assume from now on that
\bea
\label{operators}
f\left(\Box_c\right)=e^{-h\left(\Box_c\right)} \ ; \ f\left(\Box_a\right)=e^{-h\left(\Box_a\right)} \ ; \ g(\Box)=e^{-\frac{1}{2}h\left(\Box\right)} \ .
\eea
Equations (\ref{prop}) and (\ref{operators}) make clear that, through an appropriate choice of the operator $\gamma$, we can improve the UV behavior of the theory without introducing unwanted degrees of freedom. In particular, in the Fermi-Feynman gauge $\alpha=1$, the propagator takes its simplest form, which is
\bea
\label{feyngauge}
\left\langle V^A\left(-p,\theta\right)V^B\left(p,\theta^\prime\right)\right\rangle=-\frac{e^{h(-p^2)}}{p^2}\delta^{AB}\delta^4(\theta-\theta^\prime) \ .
\eea
This indicates that the theory describes only one multiplet represented by the pole $p^2=0$. 

So far, we have only considered the pure gauge theory, without matter. Now, let us introduce a coupling between the gauge superfield and the (anti)chiral matter superfields. In this paper, we assume that the interaction term between these superfields is given by \cite{GGRS}
\bea
\label{matter}
S_M[\bar\Phi,\Phi,V]=\int d^8z \bar\Phi^i{\left(e^{2gV}\right)_i}^j\Phi_j \  .
\eea
Note that we are considering here massless antichiral and chiral superfields transforming in the fundamental representation of the gauge group. It is worth to point out that the insertion of non-local operators in (\ref{matter}) n a manner compatible with the gauge symmetry would introduce very complicated interaction terms, which would make the calculations to be extremely technically difficult even in the one-loop approximation. Therefore, for the sake of simplicity, we assume that the matter action (\ref{matter}) does not contain non-local operators.

Finally, it follows from (\ref{gaugeaction}), (\ref{gaugefixing}), (\ref{operators}), and (\ref{matter}) that the complete gauge-fixed action is given by
\bea
\label{total}
S[\bar\Phi,\Phi,V]&=&\frac{1}{16g^2}\textrm{Tr}\left\{\int d^6z W^\alpha e^{-h\left(\Box_c\right)}W_\alpha+\int d^6\bar z \bar W^{\dot{\alpha}}e^{-h\left(\Box_a\right)}\bar W_{\dot{\alpha}}\right\}\nonumber\\
&&-\frac{1}{\alpha}\textrm{Tr}\int d^8z\left(e^{-\frac{1}{2}h\left(\Box\right)}D^2V\right)\left(e^{-\frac{1}{2}h\left(\Box\right)}\bar D^2V\right)+\int d^8z \bar\Phi^i{\left(e^{2gV}\right)_i}^j\Phi_j \  .
\eea
This is the non-local supersymmetric gauge theory which we will study in this paper. Notice that we have omitted the ghost term in (\ref{total}) due to the fact that the ghosts will not couple to the background chiral superfields, so that the such term does not contribute to the one-loop K\"{a}hler effective potential \cite{WGR}.   

\section{One-loop effective potential}
\label{1loop}

In this section, our goal is to calculate the one-loop correction to the K\"{a}hlerian effective potential (KEP) within the model (\ref{total}). This potential can be found through the background field method \cite{BOS}. Following this method, let us split the superfields into background $(V^A,\Phi_i,\bar\Phi^i)$ and quantum $(v^A,\phi_i,\bar\phi^i)$ parts as follows
\bea
V^A\rightarrow V^A+v^A \ ; \  \Phi_i\rightarrow\Phi_i+\phi_i \ ; \  \ \bar\Phi^i\rightarrow\bar\Phi^i+\bar\phi^i \ .
\eea
Since the KEP depends only on chiral and antichiral superfields, but not on their derivatives, we assume that the background superfields are subject to the following constraints \cite{BKY}
\bea
V^A=0 \ ; \ D_\alpha\Phi_i=0 \ ; \ \bar D_{\dot{\alpha}}\bar\Phi^i=0 \ ; \ \partial_{\alpha\dot{\alpha}}\Phi_i=0 \ ; \ \partial_{\alpha\dot{\alpha}}\bar\Phi^i=0 \ .
\eea
Thus, by expanding the action (\ref{total}) around the background superfields and keeping only the quadratic terms in the quantum superfields, we obtain
\bea
\label{quadratic}
S_2[\bar\Phi,\Phi;\bar\phi,\phi,v]&=&\frac{1}{2}\int d^8z\bigg\{ v^A\left[\delta_{AB} \ e^{-h\left(\Box\right)}\left(D^\alpha\bar D^2D_\alpha-\alpha^{-1}\left\{D^2,\bar D^2\right\}\right)+M_{AB}^2\right]v^B\nonumber\\
&&+2\bar\phi^i\phi_i+2v^A\bar X_A^i\phi_i+2\bar\phi^i X_{iA}v^A \bigg\} \  ,
\eea 
where
\bea
\bar X_A^i=2g\bar\Phi^j{\left(T_A\right)_j}^i \ ; \ X_{iA}=2g{\left(T_A\right)_i}^j\Phi_j \ ; \ M_{AB}^2=\frac{1}{2}\left(\bar X_A^iX_{iB}+\bar X_B^iX_{iA}\right) \ .
\eea
It is convenient to express the quantum antichiral and chiral superfields in terms of the superfields $\psi$ and $\bar{\psi}$, which are free of differential constraints, such that $\phi=\bar D^2\psi$ and $\bar\phi=D^2\bar\psi$ \cite{GRU}. However, this change of variables introduces a new gauge invariance under the transformations $\delta\psi=\bar D^{\dot{\alpha}}\bar\omega_{\dot{\alpha}}$ and $\delta\bar\psi=D^{\alpha}\omega_{\alpha}$. Therefore, in order to fix the gauge, we consider the following gauge-fixing functional \cite{GGRS}
\bea
\label{gaugefixing2}
S_{GF}[\psi,\bar\psi]=\int d^8z\bar\psi^i\left(\bar D^2D^2-D^\alpha\bar D^2D_\alpha\right)\psi_i \ .
\eea
For this gauge-fixing choice, the ghosts completely decouple, so that their action will be omitted. Therefore, it follows from (\ref{quadratic}) and (\ref{gaugefixing2}) that
\bea
\label{S2}
S_2[\Phi,\bar\Phi;\psi,\bar\psi,v]+S_{GF}[\psi,\bar\psi]=\frac{1}{2}\int d^8z
\left(\begin{array}{ccc}
v^A & \psi^i & \bar\psi^i
\end{array}\right)
\widehat{{\mathcal O}}
\left(\begin{array}{c}
v^B \\
\psi_j \\
\bar\psi_j
\end{array}\right) \ ,
\eea
where
\bea
\label{operator}
\widehat{{\mathcal O}}=\left(\begin{array}{ccc}
\delta_{AB} \ e^{-h\left(\Box\right)}\left(D^\alpha\bar D^2D_\alpha-\alpha^{-1}\left\{D^2,\bar D^2\right\}\right)+M_{AB}^2 \ & \ \bar X_A^j\bar D^2 \ & \ X_A^j D^2\\
\bar X_{iB}\bar D^2 \ & \ 0 \ & \ {\delta_i}^j\Box\\
X_{iB}D^2 \ & \ {\delta_i}^j\Box \ & \ 0
\end{array}\right) \ .
\eea
By formally integrating out the quantum superfields in Eq. (\ref{S2}), we arrive at the one-loop effective action, which is given by the standard general expression \cite{BOS,GRU}
\bea
\label{1loopEA}
\Gamma^{(1)}[\bar\Phi,\Phi]=-\frac{1}{2}\textrm{Tr}\ln\widehat{\mathcal O}=-\frac{1}{2}\int d^8z\textrm{tr}\ln\widehat{\mathcal O}\delta^8(z-z^{\prime})|_{z=z^\prime} \ , 
\eea
where $\textrm{tr}$ denotes the matrix trace over the internal indices.

In order to evaluate the above trace, it is convenient to use the matrix identity \cite{PP}
\bea
\label{ident}
\textrm{Tr}\ln\left(\begin{array}{cc}
\widehat{A}&\widehat{B}\\
\widehat{C} & \widehat{E}\\
\end{array}\right)=\textrm{Tr}\ln\widehat{E}+\textrm{Tr}\ln\left(\widehat{A}-\widehat{B}\widehat{E}^{-1}\widehat{C}\right) \ .
\eea
Therefore, it follows from (\ref{operator}-\ref{ident}) that
\bea
\Gamma^{(1)}[\bar\Phi,\Phi]&=&-\frac{1}{2}\textrm{Tr}\ln\left(\begin{array}{cc}
0 & {\delta_i}^j\Box\\
{\delta_i}^j\Box & 0\\
\end{array}\right)-\frac{1}{2}\textrm{Tr}\ln\bigg[\delta_{AB} \ e^{-h\left(\Box\right)}\left(D^\alpha\bar D^2D_\alpha-\alpha^{-1}\left\{D^2,\bar D^2\right\}\right)\nonumber\\
 &+&M_{AB}^2-\bar X_A^iX_{iB}\frac{\bar D^2 D^2}{\Box}-X_A^i\bar X_{iB}\frac{D^2\bar D^2}{\Box}\bigg]\\
\label{EA}
&=&-\frac{1}{2}\textrm{Tr}\ln\left(\begin{array}{cc}
0 & {\delta_i}^j\Box\\
{\delta_i}^j\Box & 0\\
\end{array}\right)-\frac{1}{2}\textrm{Tr}\ln\left[-\delta_{AB}\Box\ e^{-h\left(\Box\right)}\left(\Pi_{1/2}+\alpha^{-1}\Pi_0\right)\right]\nonumber \\
&-&\frac{1}{2}\textrm{Tr}\ln\left(\delta_{AB}-\frac{1}{\Box}e^{h\left(\Box\right)}M_{AB}^2\Pi_{1/2}+\frac{\alpha}{\Box}e^{h\left(\Box\right)}S_{AB}^2\Pi_{0+}+\frac{\alpha}{\Box}e^{h\left(\Box\right)}S_{BA}^2\Pi_{0-}\right) \ ,
\eea
where $S_{AB}^2=\frac{1}{2}\left(\bar X_A^iX_{iB}-\bar X_B^iX_{iA}\right)$ and we have introduced the projection operators \cite{West}
\bea
\Pi_{1/2}=-\frac{D^\alpha\bar D^2D_\alpha}{\Box}\ ; \ \Pi_{0+}=\frac{\bar D^2 D^2}{\Box} \ ; \ \Pi_{0-}=\frac{D^2\bar D^2}{\Box} \ ; \ \Pi_{0}=\Pi_{0+}+\Pi_{0-} \ .
\eea
Note that the first and second traces in Eq. (\ref{EA}) do not depend on the background superfields, then we can drop them out through the normalization of the effective action. Therefore, we can rewrite (\ref{EA}) as
\bea
\label{EA2}
\Gamma^{(1)}[\bar\Phi,\Phi]&=&-\frac{1}{2}\textrm{Tr}\left[\ln\left(\delta_{AB}-\frac{1}{\Box}e^{h\left(\Box\right)}M_{AB}^2\right)\Pi_{1/2}\right]-\frac{1}{2}\textrm{Tr}\left[\ln\left(\delta_{AB}+\frac{\alpha}{\Box}e^{h\left(\Box\right)}S_{AB}^2\right)\Pi_{0+}\right]\\ \nonumber
&-&\frac{1}{2}\textrm{Tr}\left[\ln\left(\delta_{AB}+\frac{\alpha}{\Box}e^{h\left(\Box\right)}S_{BA}^2\right)\Pi_{0-}\right] \ ,
\eea
where we have used the fact that the projection operators satisfy the usual relations of idempotence and orthogonality: $\Pi_{i}^2=\Pi_{i}$ and $\Pi_i\Pi_j=0$, for $i\neq j$, where $i=1/2, 0+, 0-$.

Finally, it follows from (\ref{1loopEA}) and (\ref{EA2}) that the one-loop correction to the KEP is given by
\bea
\label{genericeffecPot}
K^{(1)}(\bar\Phi,\Phi)=-\frac{1}{(4\pi)^2}\int_0^\infty dp^2\textrm{tr}\left[\ln\left(\delta_{AB}+\frac{M_{AB}^2}{p^2}e^{h\left(-p^2\right)}\right)-\ln\left(\delta_{AB}-\alpha\frac{S_{AB}^2}{p^2}e^{h\left(-p^2\right)}\right)\right] \ .
\eea 
Notice that the KEP takes its simplest form in the Landau gauge $\alpha=0$. For this reason, we will adopt the Landau gauge in our calculations from now on. On the other hand, it is worth to point out that if the gauge group is $U(1)$, therefore one has $S_{AB}^2=0$, so that the KEP is independent of the gauge parameter $\alpha$ in this particular case. We note nevertheless that it is natural to expect that the KEP will be gauge independent as it occurs in other supergauge theories.

\section{Non-local models}
\label{NL}

The last step of our calculation is to evaluate the integral over the momentum in Eq (\ref{genericeffecPot}). In order to do that, we must choose the explicit form of the entire function $h(-p^2)$. Unfortunately, it is not an easy task to choose a $h(-p^2)$ which yields an integral with exact and UV-finite solution. Thus, for simplicity, we will limit ourselves to consider two non-local operators which allow us to find approximated and UV-finite solutions for the KEP (\ref{genericeffecPot}).

\subsection{Model I}

Our first and simplest model is characterized by the entire function
\bea
\label{h1}
h_I\left(-p^2\right)=-\frac{p^2}{\Lambda^2} \ ,
\eea
where $\Lambda$ is a mass scale in the theory. For the sake of convenience, all Feynman integrals will be calculated in the dimensional regularization scheme. Thus, it follows from (\ref{genericeffecPot}) and (\ref{h1}) that 
\bea
\label{model1}
K^{(1)}_I(\bar\Phi,\Phi)=-\frac{\mu^{2\varepsilon}}{(4\pi)^{2-\varepsilon}\Gamma\left(2-\varepsilon\right)}\sum_G\int_0^\infty dp^2(p^2)^{-\varepsilon}\ln\left(1+\frac{m_G^2}{p^2}e^{-\frac{p^2}{\Lambda^2}}\right) \ ,
\eea
where $\varepsilon=2-\frac{d}{2}\rightarrow 0$, while $m_G^2$ are the eigenvalues of the mass-squared matrix $M_{AB}^2$. We note that all background dependence is concentrated just in $m_G^2$, hence they are not constants but functions of background chiral and antichiral fields, thus, their integral over the superspace does not vanish. The arbitrary mass scale $\mu$ has been introduced so that the canonical dimension of the KEP is independent of the space-time dimension $d$.

It follows from Eq. (\ref{model1}) that in the local limit $\Lambda\rightarrow\infty$, the one-loop KEP for the standard gauge theory is reproduced \cite{WGR,PW}. This suggests that if we assume the approximation $m^2_G\ll\Lambda^2$, we can obtain the deviation from the one-loop KEP obtained in the local theory. Therefore, in order to solve (\ref{model1}), we will make the assumption that $m^2_G\ll\Lambda^2$.

For the purposes of this paper, it will be appropriate to evaluate (\ref{model1}) with the help of the strategy of expansion by regions \cite{BS,Smirnov}. While this approximation method is quite useful and often used for calculations in the framework of the effective field theories \cite{BBF}, the strategy of regions was also successfully applied to the calculation of the one-loop KEP within the context of non-local chiral superfield theories in \cite{BGNP}.

Following the method, let us introduce an intermediate scale $\Omega^2$, such that $m^2_G\ll\Omega^2\ll\Lambda^2$. This allows us to split the interval of integration into two regions, which are called the low-energy region $[0,\Omega^2]$ and the high-energy region $[\Omega^2,\infty)$. Therefore, we can rewrite the integral (\ref{model1}) as
\bea
\label{split1}
K^{(1)}_{I}(\bar\Phi,\Phi)&=&-\frac{1}{16\pi^2}\frac{(4\pi\mu^2)^\varepsilon}{\Gamma\left(2-\varepsilon\right)}\sum_G\left[\int_0^{\Omega^2} dp^2+\int_{\Omega^2}^\infty dp^2\right](p^2)^{-\varepsilon}\ln\left(1+\frac{m_G^2}{p^2}e^{-\frac{p^2}{\Lambda^2}}\right)\nonumber\\
&\equiv&-\frac{1}{16\pi^2}\sum_G\left[I_L+I_H\right] \ .
\eea 
Due to the fact that $p^2\sim m^2_G\ll\Lambda^2$ in the low-energy region $[0,\Omega^2]$, the integral $I_L$ can be expanded in powers of $1/\Lambda^2$. Thus, we obtain
\bea
\label{I_L-1}
I_L=\frac{(4\pi\mu^2)^\varepsilon}{\Gamma\left(2-\varepsilon\right)}\int_0^{\Omega^2} dp^2(p^2)^{-\varepsilon}\Bigg\{\ln\left(1+\frac{m_G^2}{p^2}\right)-\frac{m_G^2}{\Lambda^2}\frac{p^2}{p^2+m_G^2}+\frac{m_G^2}{2\Lambda^4}\frac{p^6}{\big(p^2+m_G^2\big)^2}+\cdots\Bigg\} \ ,
\eea
where we have kept terms up to the second order in $1/\Lambda^2$.

On the other hand, since $m^2_G\ll p^2\sim\Lambda^2$ in the high-energy region $[\Omega^2,\infty)$, we can expand $I_H$ in powers of $m^2_G$ and keep terms up to the third order in $m^2_G$. Therefore, we get
\bea
\label{I_H-1}
I_H=\frac{(4\pi\mu^2)^\varepsilon}{\Gamma\left(2-\varepsilon\right)}\int_{\Omega^2}^\infty dp^2(p^2)^{-\varepsilon}\Bigg\{m_G^2\frac{e^{-\frac{p^2}{\Lambda^2}}}{p^2}-\frac{m_G^4}{2}\frac{e^{-2\frac{p^2}{\Lambda^2}}}{p^4}+\frac{m_G^6}{3}\frac{e^{-3\frac{p^2}{\Lambda^2}}}{p^6}+\cdots\Bigg\} \ .
\eea
We can simplify our calculations by splitting the integrals (\ref{I_L-1}-\ref{I_H-1}) into
\bea
\label{SI_L-1}
I_L&=&\frac{(4\pi\mu^2)^{\varepsilon}}{\Gamma(2-\varepsilon)}\left[\int_{0}^\infty dp^2-\int_{\Omega^2}^\infty dp^2\right](p^2)^{-\varepsilon}\bigg\{\ln\left(1+\frac{m_G^2}{p^2}\right)-\frac{m_G^2}{\Lambda^2}\frac{p^2}{p^2+m_G^2}\nonumber\\
&+&\frac{m_G^2}{2\Lambda^4}\frac{p^6}{\big(p^2+m_G^2\big)^2}+\cdots\bigg\}\equiv I_L\big|_{\Omega^2\rightarrow\infty}-R_L \ ,
\eea
and
\bea
\label{SI_H-1}
I_H&=&\frac{(4\pi\mu^2)^{\varepsilon}}{\Gamma(2-\varepsilon)}\left[\int_{0}^\infty dp^2-\int_{0}^{\Omega^2} dp^2\right](p^2)^{-\varepsilon}\bigg\{m_G^2\frac{e^{-\frac{p^2}{\Lambda^2}}}{p^2}-\frac{m_G^4}{2}\frac{e^{-2\frac{p^2}{\Lambda^2}}}{p^4}+\frac{m_G^6}{3}\frac{e^{-3\frac{p^2}{\Lambda^2}}}{p^6}+\cdots\bigg\}\nonumber\\
&\equiv& I_H\big|_{\Omega^2\rightarrow 0}-R_H \ .
\eea
The integrals in $I_L\big|_{\Omega^2\rightarrow\infty}$ and $I_H\big|_{\Omega^2\rightarrow 0}$ can be exactly evaluated. Therefore, in the limit $\varepsilon\rightarrow 0$, we find
\bea
\label{LI_L-1}
I_L\big|_{\Omega^2\rightarrow\infty}&=&m^2_G\left[\frac{1}{\varepsilon}+2-\gamma-\ln\left(\frac{m_G^2}{4\pi\mu^2}\right)\right]+\frac{m^4_G}{\Lambda^2}\bigg[\frac{1}{\varepsilon}+1-\gamma-\ln\left(\frac{m^2_G}{4\pi\mu^2}\right)\bigg]\nonumber\\
&+&\frac{m^6_G}{2\Lambda^4}\bigg[\frac{3}{\varepsilon}+2-3\gamma-3\ln\left(\frac{m^2_G}{4\pi\mu^2}\right)\bigg] \ ; \\
\label{LI_H-1}
I_H\big|_{\Omega^2\rightarrow 0}&=&-m_G^2\left[\frac{1}{\varepsilon}+1+\ln\left(\frac{4\pi\mu^2}{\Lambda^2}\right)\right]-\frac{m^4_G}{\Lambda^2}\left[\frac{1}{\varepsilon}+\ln\left(\frac{8\pi\mu^2}{\Lambda^2}\right)\right]-\frac{3m_G^6}{4\Lambda^4}\nonumber\\
&\times&\left[\frac{2}{\varepsilon}-1+2\ln\left(\frac{12\pi\mu^2}{\Lambda^2}\right)\right] \ .
\eea
We notice that the divergences in (\ref{LI_L-1}) and (\ref{LI_H-1}) are of ultraviolet and infrared nature, respectively. Moreover, it is worth pointing out that they are spurious divergences which will cancel out in the final result for the KEP.  

Notice that $R_L$ and $R_H$ depend on the artificial scale $\Omega^2$. However, since $\Omega^2$ is not present in the integral (\ref{model1}), this implies that $R_L+R_H$ must be independent of $\Omega^2$. In order to prove this, let us expand $R_L$ in powers of $m_G^2$ and keep terms up to the third order in $m^2_G$. Therefore, we have
\bea
\label{R_L-1}
R_L&=&\frac{(4\pi\mu^2)^{\varepsilon}}{\Gamma(2-\varepsilon)}\int_{\Omega^2}^\infty dp^2(p^2)^{-\varepsilon}\bigg\{\left(-\frac{1}{\Lambda^2}+\frac{1}{p^2}+\frac{p^2}{2\Lambda^4}\right)m_G^2+\left(-\frac{1}{\Lambda^4}-\frac{1}{2p^4}+\frac{1}{\Lambda^2 p^2}\right)m_G^4\nonumber\\
&+&\left(\frac{1}{3p^6}-\frac{1}{\Lambda^2 p^4}+\frac{3}{2\Lambda^4 p^2}\right)m_G^6+\cdots\bigg\} \ .
\eea
On the other hand, we can expand $R_H$ in powers of $1/\Lambda^2$ and keep terms up to second order in $1/\Lambda^2$. Thus, we find
\bea
\label{R_H-1}
R_H&=&\frac{(4\pi\mu^2)^{\varepsilon}}{\Gamma(2-\varepsilon)}\int_{0}^{\Omega^2} dp^2(p^2)^{-\varepsilon}\bigg\{\left(\frac{m_G^6}{3p^6}-\frac{m_G^4}{2p^4}+\frac{m_G^2}{p^2}\right)+\left(-m_G^2+\frac{m_G^4}{p^2}-\frac{m^6_G}{p^4}\right)\frac{1}{\Lambda^2}\nonumber\\
&+&\left(-m_G^4+\frac{3m_G^6}{2p^2}+\frac{m^2_G p^2}{2}\right)\frac{1}{\Lambda^4}+\cdots\bigg\} \ .
\eea
It is not necessary to explicitly evaluate the integrals in $R_L$ and $R_H$. Since the integrands of $R_L$ and $R_H$ are equal, the sum of (\ref{R_L-1}) and {\ref{R_H-1} is given by
\bea
\label{R_L+R_H}
R_L+R_H&=&\frac{(4\pi\mu^2)^{\varepsilon}}{\Gamma(2-\varepsilon)}\int_{0}^{\infty} dp^2(p^2)^{-\varepsilon}\bigg\{\left(\frac{m_G^6}{3p^6}-\frac{m_G^4}{2p^4}+\frac{m_G^2}{p^2}\right)+\left(-m_G^2+\frac{m_G^4}{p^2}-\frac{m^6_G}{p^4}\right)\frac{1}{\Lambda^2}\nonumber\\
&+&\left(-m_G^4+\frac{3m_G^6}{2p^2}+\frac{m^2_G p^2}{2}\right)\frac{1}{\Lambda^4}+\cdots\bigg\} \ .
\eea
Note that the above integrals are evaluated over the full integration interval $p^2\in[0,\infty)$, so that $R_L+R_H$ is independent of $\Omega^2$, what was to be demonstrated. Moreover, the vanishing of these integrals within the dimensional regularization scheme implies that $R_L+R_H=0$.

Finally, since $I_L+I_H=I_L\big|_{\Omega^2\rightarrow\infty}+I_H\big|_{\Omega^2\rightarrow 0}$, we can substitute (\ref{LI_L-1}) and (\ref{LI_H-1}) into (\ref{split1}) to get
\bea
\label{kep1}
K^{(1)}_{I}(\bar\Phi,\Phi)&\approx&\frac{1}{16\pi^2}\sum_G\bigg\{m_G^2\ln\left(\frac{m_G^2}{e^{1-\gamma}\Lambda^2}\right)+\frac{m_G^4}{\Lambda^2}\ln\left(\frac{2m_G^2}{e^{1-\gamma}\Lambda^2}\right)+\frac{m_G^6}{4\Lambda^4}\bigg[-1+6\nonumber\\
&\times&\ln\left(\frac{3m_G^2}{e^{1-\gamma}\Lambda^2}\right)\bigg]\bigg\} \ .
\eea
We notice that the one-loop correction for the KEP is finite. As we have already anticipated, the spurious divergences were completely cancelled in the sum of (\ref{LI_L-1}) and (\ref{LI_H-1}). Furthermore, we also notice that the final result (\ref{kep1}) is independent of $\mu$, so that the KEP is scale invariant. Lastly, in the limit $\Lambda\rightarrow\infty$, the result (\ref{kep1}) coincides with the one-loop KEP obtained in \cite{WGR} for the local gauge theory, which ensures the consistency of our result.

\subsection{Model II}

Our second model is described by the fourth-degree polynomial
\bea
\label{h2}
h_{II}\left(-p^2\right)=-\frac{1}{\Lambda^2}\left(p^2+\frac{p^4}{\Sigma^2}\right) \ ,
\eea
where $\Lambda$ and $\Sigma$ are mass scales in which the non-local contributions become relevant. Note that the previous model (\ref{h1}) is a particular case of (\ref{h2}) when $\Sigma\rightarrow\infty$. Thus, the operator $h_{II}(\Box)$ can be thought as a higher-derivative generalization of $h_{I}(\Box)$.

Thus, it follows from (\ref{genericeffecPot}) and (\ref{h2}) that the dimensionally regularized KEP is given by 
\bea
\label{model2}
K^{(1)}_{II}(\bar\Phi,\Phi)=-\frac{\mu^{2\varepsilon}}{(4\pi)^{2-\varepsilon}\Gamma\left(2-\varepsilon\right)}\sum_G\int_0^\infty dp^2(p^2)^{-\varepsilon}\ln\left\{1+\frac{m_G^2}{p^2}\exp\left[-\frac{1}{\Lambda^2}\left(p^2+\frac{p^4}{\Sigma^2}\right)\right] \right\} \ .
\eea
Again, we will apply the strategy of expansion by regions to obtain  an explicit approximate solution for (\ref{model2}). In particular, we assume that $m_G^2\ll\Lambda^2,\Sigma^2$, so that $m_G^2\ll\Omega^2\ll\Lambda^2,\Sigma^2$. Thus, we can split the integral (\ref{model2}) into the following low- and high-energy contributions
\bea
\label{split2}
K^{(1)}_{II}(\bar\Phi,\Phi)&=&-\frac{1}{16\pi^2}\frac{(4\pi\mu^2)^\varepsilon}{\Gamma\left(2-\varepsilon\right)}\sum_G\left[\int_0^{\Omega^2} dp^2+\int_{\Omega^2}^\infty dp^2\right](p^2)^{-\varepsilon}\nonumber\\
&\times&\ln\left\{1+\frac{m_G^2}{p^2}\exp\left[-\frac{1}{\Lambda^2}\left(p^2+\frac{p^4}{\Sigma^2}\right)\right]\right\}\equiv-\frac{1}{16\pi^2}\sum_G\left[I_L+I_H\right] \ .
\eea 
On the one hand, we notice that $p^2\sim m^2_G\ll\Lambda^2,\Sigma^2$ in the low-energy region $[0,\Omega^2]$. Thus, it follows that we can expand the integral $I_L$ in powers of $1/\Lambda^2$ and $1/\Sigma^2$ to get
\bea
\label{I_L-2}
I_L&=&\frac{(4\pi\mu^2)^\varepsilon}{\Gamma\left(2-\varepsilon\right)}\int_0^{\Omega^2} dp^2(p^2)^{-\varepsilon}\Bigg\{\ln\left(1+\frac{m_G^2}{p^2}\right)-\frac{m_G^2}{\Lambda^2}\frac{p^2}{p^2+m_G^2}+\frac{m_G^2}{2\Lambda^4}\frac{p^6}{\big(p^2+m_G^2\big)^2}\nonumber\\
&-&\frac{m_G^2}{\Lambda^2\Sigma^2}\frac{p^4}{p^2+m^2_G}+\cdots\Bigg\} \ ,
\eea
where we have kept terms up to the second order in $1/\Lambda^2$ and $1/\Sigma^2$.

On the other hand, we also notice that $m^2_G\ll p^2\sim\Lambda^2$ in the high-energy region $[\Omega^2,\infty)$. Thus, $I_H$ can be expanded in powers of $m^2_G$ up to the third order in $m^2_G$. Therefore, we obtain
\bea
\label{I_H-2}
I_H&=&\frac{(4\pi\mu^2)^\varepsilon}{\Gamma\left(2-\varepsilon\right)}\int_{\Omega^2}^\infty dp^2(p^2)^{-\varepsilon}\Bigg\{\frac{m_G^2}{p^2}\exp\left[-\frac{1}{\Lambda^2}\left(p^2+\frac{p^4}{\Sigma^2}\right)\right]-\frac{m_G^4}{2p^4}\exp\left[-\frac{2}{\Lambda^2}\left(p^2+\frac{p^4}{\Sigma^2}\right)\right]\nonumber\\
&+&\frac{m_G^6}{3p^6}\exp\left[-\frac{3}{\Lambda^2}\left(p^2+\frac{p^4}{\Sigma^2}\right)\right]+\cdots\Bigg\} \ .
\eea
At this point of the calculation, we can repeat the same argument from the previous subsection to prove that $I_L+I_H=I_L\big|_{\Omega^2\rightarrow\infty}+I_H\big|_{\Omega^2\rightarrow 0}$. Thus, it is only necessary to evaluate the integrals in $I_L\big|_{\Omega^2\rightarrow\infty}$ and $I_H\big|_{\Omega^2\rightarrow 0}$. Therefore, in the limit $\varepsilon\rightarrow 0$, we find
\bea
\label{LI_L-2}
I_L\big|_{\Omega^2\rightarrow\infty}&=&m^2_G\left[\frac{1}{\varepsilon}+2-\gamma-\ln\left(\frac{m_G^2}{4\pi\mu^2}\right)\right]+\frac{m^4_G}{\Lambda^2}\bigg[\frac{1}{\varepsilon}+1-\gamma-\ln\left(\frac{m^2_G}{4\pi\mu^2}\right)\bigg]\nonumber\\
&+&\frac{m^6_G}{2\Lambda^4}\bigg[\frac{3}{\varepsilon}+2-3\gamma-3\ln\left(\frac{m^2_G}{4\pi\mu^2}\right)\bigg]+\frac{m^6_G}{\Lambda^2\Sigma^2}\bigg[-\frac{1}{\varepsilon}-1+\gamma+\ln\left(\frac{m^2_G}{4\pi\mu^2}\right)\bigg] \ ; \\
\label{LI_H-2}
I_H\big|_{\Omega^2\rightarrow 0}&=&-m_G^2\left[\frac{1}{\varepsilon}+1+\ln\left(\frac{8\pi\mu^2}{\Lambda\Sigma}\right)-\frac{1}{2}U^{(1,0,0)}\left(0,\frac{1}{2},\frac{\Sigma^2}{4\Lambda^2}\right)\right]-\frac{m^4_G}{\Lambda^2}\Bigg[\frac{1}{\varepsilon}\nonumber\\
&+&\ln\left(\frac{8\sqrt{2}\pi\mu^2}{\Lambda\Sigma}\right)-\frac{1}{2}U^{(1,0,0)}\left(0,\frac{3}{2},\frac{\Sigma^2}{2\Lambda^2}\right)\Bigg]-\frac{3m_G^6}{4\Lambda^4}\left[\frac{2}{\varepsilon}-1+2\ln\left(\frac{8\sqrt{3}\pi\mu^2}{\Lambda\Sigma}\right)\right]\nonumber\\
&+&\frac{m_G^6}{\Lambda^2\Sigma^2}\left[\frac{1}{\varepsilon}-\frac{1}{2}+\ln\left(\frac{8\sqrt{3}\pi\mu^2}{\Lambda\Sigma}\right)+U^{(1,0,0)}\left(-1,\frac{1}{2},\frac{3\Sigma^2}{4\Lambda^2}\right)\right] \ ,
\eea
where $U^{(1,0,0)}(a,b,z)$ is defined in terms of the confluent hypergeometric function of the second kind $U(a,b,z)$ by \cite{Weisstein}:
\bea
U^{(1,0,0)}(a,b,z)\equiv\frac{\partial}{\partial d}U(d,b,z)\bigg|_{d=a} \ .
\eea
By substituting (\ref{LI_L-2}) and (\ref{LI_H-2}) into (\ref{split2}) we find
\bea
\label{kep2}
K^{(1)}_{II}(\bar\Phi,\Phi)&\approx&\frac{1}{16\pi^2}\sum_G\Bigg\{m_G^2\left[\ln\left(\frac{2m_G^2}{e^{1-\gamma}\Lambda\Sigma}\right)-\frac{1}{2}U^{(1,0,0)}\left(0,\frac{1}{2},\frac{\Sigma^2}{4\Lambda^2}\right)\right]+\frac{m_G^4}{\Lambda^2}\bigg[\ln\left(\frac{2\sqrt{2}m_G^2}{e^{1-\gamma}\Lambda\Sigma}\right)\nonumber\\
&-&\frac{1}{2}U^{(1,0,0)}\left(0,\frac{3}{2},\frac{\Sigma^2}{2\Lambda^2}\right)\bigg]+\frac{m_G^6}{4\Lambda^4}\left[-1+6\ln\left(\frac{2\sqrt{3}m_G^2}{e^{1-\gamma}\Lambda\Sigma}\right)\right]-\frac{m_G^6}{\Lambda^2\Sigma^2}\bigg[\ln\left(\frac{2\sqrt{3}m_G^2}{e^{\frac{3}{2}-\gamma}\Lambda\Sigma}\right)\nonumber\\
&+&U^{(1,0,0)}\left(-1,\frac{1}{2},\frac{3\Sigma^2}{4\Lambda^2}\right)\bigg]\Bigg\} \ .
\eea
Just like the previous model, the spurious divergences and arbitrary scale $\mu$ were completely cancelled in the sum of (\ref{LI_L-2}) and (\ref{LI_H-2}) to produce a finite and scale invariant KEP (\ref{kep2}). 

In order to check if the one-loop KEP (\ref{kep2}) reproduces the one of the previous model (\ref{kep1}) in the low-energy limit $\Sigma\rightarrow\infty$, we need to determine the asymptotic form of $K^{(1)}_{II}(\bar\Phi,\Phi)$ for $\Sigma\gg\Lambda$. This result can be obtained by substituting the following  asymptotic expansions:  
\bea
U^{(1,0,0)}\left(0,\frac{1}{2},\frac{\Sigma^2}{4\Lambda^2}\right)&\approx&-2\ln\left(\frac{\Sigma}{2\Lambda}\right)-\frac{2\Lambda^2}{\Sigma^2}+\frac{6\Lambda^4}{\Sigma^4}+\mathcal{O}\left(\frac{\Lambda^6}{\Sigma^6}\right) \ ;\nonumber\\
U^{(1,0,0)}\left(0,\frac{3}{2},\frac{\Sigma^2}{2\Lambda^2}\right)&\approx&-2\ln\left(\frac{\Sigma}{\sqrt{2}\Lambda}\right)+\frac{\Lambda^2}{\Sigma^2}-\frac{\Lambda^4}{2\Sigma^4}+\mathcal{O}\left(\frac{\Lambda^6}{\Sigma^6}\right) \ ;\nonumber\\
\frac{4\Lambda^2}{3\Sigma^2}U^{(1,0,0)}\left(-1,\frac{1}{2},\frac{3\Sigma^2}{4\Lambda^2}\right)&\approx&\left(\frac{4\Lambda^2}{3\Sigma^2}-2\right)\ln\left(\frac{\sqrt{3}\Sigma}{2\Lambda}\right)+\frac{2\Lambda^2}{\Sigma^2}+\frac{2\Lambda^4}{9\Sigma^4}+\mathcal{O}\left(\frac{\Lambda^6}{\Sigma^6}\right)
\eea
into Eq. (\ref{kep2}). Thus, it is trivial to show that
\bea
\label{kep22}
K^{(1)}_{II}(\bar\Phi,\Phi)&\approx&K^{(1)}_{I}(\bar\Phi,\Phi)+\frac{1}{16\pi^2}\sum_G\bigg\{m_G^2\left[\frac{\Lambda^2}{\Sigma^2}-\frac{3\Lambda^4}{\Sigma^4}+\mathcal{O}\left(\frac{\Lambda^6}{\Sigma^6}\right)\right]+\frac{m_G^4}{\Lambda^2}\bigg[-\frac{\Lambda^2}{2\Sigma^2}+\frac{\Lambda^4}{4\Sigma^4}\nonumber\\&+&\mathcal{O}\left(\frac{\Lambda^6}{\Sigma^6}\right)\bigg]+\frac{m_G^6}{\Lambda^4}\bigg[-\frac{\Lambda^2}{\Sigma^2}\ln\left(\frac{3m_G^2}{e^{-\gamma}\Lambda^2}\right)-\frac{\Lambda^4}{6\Sigma^4}+\mathcal{O}\left(\frac{\Lambda^6}{\Sigma^6}\right)\bigg]\bigg\} \ .
\eea
From this asymptotic form of $K^{(1)}_{II}(\bar\Phi,\Phi)$, it is manifest that the physical effects of the higher-derivative term in $h_{II}(\Box)$ (see Eq. \ref{h2}) are suppressed for large $\Sigma$, so that all higher-derivative effects are completely decoupled in the limit $\Sigma\rightarrow\infty$.

\section{Summary}
\label{Conc}

In this paper, we formulated the nonlocal extension of the ${\cal N}=1$ super-Yang-Mills theory with the chiral matter. All our studies were based on superfield approach allowing to maintain explicit supersymmetry on all steps of calculations. We explicitly calculated the one-loop K\"{a}hlerian effective potential for two kinds of the form factor, with the gauge group can be Abelian as well as non-Abelian. Our results turn out to be explicitly UV finite, as it should be, and similar to the one-loop K\"{a}hlerian effective potential in the super-Yang-Mills theory \cite{WGR,PW}, with the role of the normalization parameter $\mu$ is played by the nonlocality scale $\Lambda$. This fact can be naturally explained since the nonlocal extension in this theory effective plays the role of the specific regularization. 

In this context, the very important problem is a study of the one-loop low-energy effective action in the non-local extension of the ${\cal N}=4$ super-Yang-Mills theory. Indeed, from one side, the nonlocality naturally improves the UV behavior, from another side, nonlocal extensions are always characterized by a some energy scale, thus, the nonlocal extension of this theory naturally will break the superconformal symmetry. due to arising of the characteristic energy scale. We plan to study this problem in a forthcoming paper.

{\bf Acknowledgements.} This work was partially supported by Conselho Nacional de Desenvolvimento Cient\'{i}­fico e Tecnol\'{o}gico (CNPq). The work by A. Yu. P. has been partially supported by the CNPq project No. 303783/2015-0.

\end{document}